# Magnetic Structures of High Temperature Phases of TbBaCo$_2$O$_{5.5}$

Minoru Soda[1], Yukio Yasui[1], Toshiaki Fujita[1], Takeshi Miyashita[1],
Masatoshi Sato[1] and Kazuhisa Kakurai[2]

[1]*Department of Physics, Division of Material Science, Nagoya University,
Furo-cho, Chikusa-ku, Nagoya 464-8602*
[2]*Advanced Science Research Center, Japan Atomic Energy Research Institute,
Tokai-mura, Naka-gun, Ibaraki 319-1195*

**Abstract**

Neutron diffraction studies have been carried out on a single crystal of oxygen-deficient perovskite TbBaCo$_2$O$_{5.5}$ in the temperature range of 7-370 K. There have been observed several magnetic or structural transitions. Among these, the existence of the transitions to the insulating phase from the metallic one at ~340 K, to the one with the ferromagnetic moment at ~280 K and possibly to the antiferromagnetic one at ~260 K, with decreasing temperature $T$ correspond to those reported in former works. We have studied the magnetic structures at 270 K and 250 K and found that all Co$^{3+}$ ions of the CoO$_6$ octahedra are in the low spin state and those of the CoO$_5$ pyramids carry spins which are possibly in the intermediate spin state. Non-collinear magnetic structures are proposed at these temperatures. Two other transitions have also been observed at the temperatures, ~100 K and ~250 K.

corresponding author: M. Sato (e-mail: msato@b-lab.phys.nagoya-u.ac.jp)

Keywords: TbBaCo$_2$O$_{5.5}$, neutron diffraction, magnetic structure, phase transitions

## 1. Introduction

In the perovskite oxides $RCoO_3$ (R=Y and rare earth elements), the $Co^{3+}$ ions have the low spin (LS; $t_{2g}^6$; $S=0$) ground state and often exhibit the spin state change[1-6] to the intermediate spin (IS; $t_{2g}^5 e_g^1$; $S=1$) state or the high spin (HS; $t_{2g}^4 e_g^2$; $S=2$) state with increasing temperature $T$, indicating that the energy difference $\delta E$ between the spin states is small. Because the energy splitting due to the crystal field generally increases with decreasing volume of the $CoO_6$ octahedra, $\delta E$ can be controlled by the volume change of the octahedra and therefore we can expect variety of physical behavior in the system related to the spin state change.

The system $R_{1-x}A_xCoO_{3-\delta}$ obtained by the substitution of R with divalent atoms A=Sr, Ba and Ca often exhibits the ferromagnetic and metallic properties for small $\delta$ with increasing doping concentration $x$ due to the double exchange interaction.[7-10] When Ba and relatively small trivalent elements are chosen as A and R, respectively, the system forms the oxygen-deficient perovskite structure, in which the oxygen deficiency is ordered.[11-18] Among such oxides, $RBaCo_2O_5$ ($\delta\sim0.5$) (R=Y, Tb, Dy and Ho) exhibits transitions due to the antiferromagnetic ordering of the Co-moments and the ordering of $Co^{2+}$ and $Co^{3+}$.[11-13] $RBaCo_2O_{5.5}$ ($\delta\sim0.25$) (R=Y, Sm, Eu, Gd, Tb, Dy and Ho) also exhibits several transitions.[14-18] However, only the restricted information is available on both the structural and the magnetic properties.

In the present work, neutron scattering studies have been carried out on a single crystal of $TbBaCo_2O_{5.5}$ to collect basic information on the spin state of Co ions and the magnetic structures. The oxygen deficient perovskite structure of this system is formed by the linkage of alternating $CoO_6$ octahedra and $CoO_5$ pyramids along the $b$-axis, as shown schematically in Fig. 1.[17,18] It has been reported that it has the orthorhombic unit cell, where the Ba and Tb are in different $c$-planes. The unit cell volume is described by $\sim a_p \times 2a_p \times 2a_p$, $a_p$ being the lattice parameter of the cubic perovskite cell. In this compound, the metal to insulator (MI) transition takes place with decreasing $T$ at $T_{MI}\sim340$ K,[14,15] as shown in the inset of Fig. 2 by the resistivity($\rho$) data taken for a sintered sample. The magnetizations $M$ measured for a sintered sample and single crystal ones prepared in the present work are shown against $T$ in the main panel of Fig. 2. As is described in detail later, the $T$-dependence of $M$ indicates at least two magnetic transitions: The ferromagnetic transition occurs at $T_C\sim280$ K, and the ferromagnetic phase suddenly changes to antiferromagnetic one at $T_N\sim260$ K with decreasing $T$ in zero magnetic field $H$ as was confirmed by neutron diffraction study ($T_N$ seems to be slightly $H$-dependent.[16]). Although transitions have been found at lower temperatures, too, for the crystals used in the present neutron diffraction studies, we mainly report here results of the studies on the magnetic structures at 270 K and 250 K, where the spin states of Co ions within the $CoO_6$ octahedra and the $CoO_5$ pyramids are basically clarified.

## 2. Experiments

A single crystal of TbBaCo$_2$O$_{5.5}$ was grown by a floating zone (FZ) method. Tb$_4$O$_7$, BaCO$_3$ and Co$_3$O$_4$ were mixed with the proper molar ratios and the mixtures were pressed into rods and calcined at 1150°C for 10 h in flowing oxygen and cooled at a rate of 100 K/h. The single crystal was grown with use of an image furnace. The obtained crystal was annealed at 1000°C for 120 h in flowing oxygen and cooled at a rate of 100 K/h. The δ value of the sample was determined by the thermo gravimetric analysis (TGA) to be δ~0.24±0.02 for the annealed one.

The magnetizations *M* were measured by using a Quantum Design SQUID magnetometer in the temperature range of 5-350 K. The *T*-dependence of *M* taken at *H*=1 T and the *H*-dependence of *M* taken at several fixed *T* by using edge parts of the crystal after the zero field cooling are shown in Figs. 2 and 3, respectively. In taking the data of *M*, the crystal orientation with respect to the ***H***-direction was so chosen that *M* at 270 K had the maximum value with the orientation-change within the condition of ***H***⊥(*c*-axis). The *T*-dependence of *M* of the as-cast crystal indicates that the ferromagnetic-antiferromagnetic transition at ~260 K is much broader than that of the polycrystal sample, while the transition of the annealed crystal is only slightly broader than that of the polycrystal sample. Then, we have used the annealed crystal in the neutron scattering studies. It is noted here that the anisotropy of the measured magnetization indicates that the ferromagnetic magnetization is within the *c*-plane.

Neutron measurements were carried out by using the triple axis spectrometer TAS-2 installed at the thermal guide of JRR-3M of JAERI in Tokai. The crystal was oriented with the [100] (or [010]) axis vertical, where both (0,*k*,*l*) and (*h*,0,*l*) points in the reciprocal space could be observed due to the coexistence of the ***a***\*- and ***b***\*-domains. Several 0*kl* reflections can be distinguished from *h*0*l* reflections by the difference between the lattice parameters *a* and *b*/2 (The parameter *a* is slightly smaller than *b*/2.). The 002 reflection of Pyrolytic graphite (PG) was used for both the monochromator and the analyzer. The horizontal collimations usually used were 17'(effective)-40'-80' (2-axis) and the neutron wavelength was ~2.44Å. A PG filter was placed in front of the sample to eliminate the higher order contamination. To realize the better resolution, horizontal collimations 17'(effective)-20'-20'-40' were also used in some cases. Another PG filter was placed after the sample in measuring reflections for which the contamination of the higher order beam may not be negligible. The sample was set in an Al-can filled with exchange He gas, which was attached to the cold head of the Displex type refrigerator for the measurements below room temperature. The can was heated in a furnace for the measurements above room temperature.

The volume ratio of the ***a***\*- and ***b***\*-domains was estimated at room temperature from the ratios of the intensities of the 0k0 reflections with odd k values to those for 0k0/h00 with even k and corresponding h values. (In the measurements of the intensities, the crystal was oriented with the [001] axis vertical.)

## 3. Experimental Results and Discussion

The intensities of the neutron Bragg scattering have been measured on the single crystal of TbBaCo$_2$O$_{5.5}$ at various $Q$-points in the reciprocal space in the temperature range of 7-370 K, where several superlattice reflections as well as the fundamental reflections of the cell with the size of $\sim a_p \times 2a_p \times 2a_p$ have been found. The points at which the reflections were observed at 7 K in the reciprocal space are shown in Fig. 4(a). ($Q=(0,k,l)$ may be ($k/2,0,l$).) The temperature dependences of the peak intensities of the typical fundamental and superlattice reflections are shown in Figs. 4(b) and 4(c), respectively. With decreasing $T$ from the temperature above 350 K, the intensities of several nuclear Bragg reflections change at $T_{MI} \sim 340$ K, and at $T_C \sim 280$ K, the additional reflection component appears at several nuclear Bragg points. It is considered to be magnetic, as is confirmed later by detailed analyses and corresponds to the appearance of the ferromagnetic moment. With further decreasing $T$, at $T_N \sim 260$ K, the additional component suddenly disappears and the superlattice peaks corresponding to the period of $2c$ ($\sim 4a_p$) along the $c$-axis appear ($Q=(0,k,1/2)$, $(0,k,3/2)$, etc.). At $T \sim 100$ K, the superlattice peaks corresponding to the period of $4c$ along the $c$-axis appear ($Q=(0,0,1 \pm 1/4)$, $(0,0,3 \pm 1/4)$, $(0,2,1 \pm 1/4)$, etc.) with decreasing $T$. The former group of the superlattice reflections may be magnetic, while we cannot distinguish if the latter superlattice reflections are magnetic or not. It should be also noted that the superlattice reflections corresponding to the modulation vector $q = a^*/3$ or $2b^*/3$ (the period of $\sim 3a_p$ in the real space) or $q=a^*/6$ or $b^*/3$ (the period of $\sim 6a_p$ in the real space) are observed even at temperatures higher than $T_{MI}$. (We could not distinguish along which direction of $a^*$ or $b^*$ the modulation exists.) We have examined the possibility that these superlattice reflections of the $\sim 3a_p$ (or $\sim 6a_p$) period originate from the coexistence of the tetragonal unit cell with the size of $\sim 3a_p \times 3a_p \times 2a_p$ reported in ref. 15 as the distinct phase with $\delta \sim 0.28$. However, by the detailed inspection of the X-ray diffraction patterns taken for the powder sample obtained by crushing a part of the crystal can exclude the possibility, because the volume of the tetragonal part is too small, if any, to explain the observed intensities of the superlattice reflections. These reflections of the $\sim 3a_p$ (or $\sim 6a_p$) period are considered to be due to the existence of the modulation of the crystal structure of the main phase in the present crystal. However, we have not clarified detailed pattern of the modulation in the present study.

The data of Fig. 4(b) suggest the existence of structural transitions in the present system at three temperatures, $T_{MI}$, $\sim 250$ K and $\sim 100$ K. The transition at $T_{MI}$ has been reported by Kusuya et al.[17] If the increase of the intensity of 020 reflection observed with decreasing $T$ below ~250 K, is due to the increase of the magnetic reflection, the system should have the net moment, because (0,2,0) corresponds to the ferromagnetic Bragg point in the reciprocal space. However, it is in contradiction to the present observation (see Fig.2). Then, the transition can be considered to be due to the structural one.

As for the crystal structure above $T_{MI}$~340 K, there exist several reports,[17,18] where the unit cell size is described by ~$a_p \times 2a_p \times 2a_p$. Of course, the existence of the ~$3a_p$ (or ~$6a_p$) period modulation along *a*- or *b*-axis observed in the present neutron diffraction measurements on the single crystal, cannot be explained. At this moment, we do not distinguish if the modulation is just a characteristic of the present single crystal sample or it exists in sintered samples, too. (Single crystals prepared from the molten phase may have different characteristics of the structure from those of sintered samples, because lattice imperfections such as the oxygen vacancies and inhomogeneity may be introduced in the course of the crystal growth.)

For the precise analyses of the magnetic structures of the present system, its crystal structure has to be first determined. It is, however, not easy due to the presence of the ~$3a_p$ (or ~$6a_p$) period modulation. Here, we simply neglect the modulation by considering that the intensities of the corresponding superlattice reflections are much weaker than the fundamental ones (less than ~ 0.5 % of the strongest one) and the crystal structure is analyzed by using the structure with the unit cell size of ~$a_p \times 2a_p \times 2a_p$ (space group P*mmm*[17]). The results of the fitting of the calculated intensities ($I_{cal}$) of the nuclear Bragg reflections to the integrated intensities ($I_{obs}$) measured at 300 K are shown in Fig. 5, where $I_{obs}$ values are plotted against $I_{cal}$. (In the fitting, we have considered the volume ratio of the ***a****\*- and ***b****\*-domains determined as stated in **2**. The consideration of this domain distribution was also made in all the analyses described below.) The lattice parameters *a*, *b* and *c* are 3.8679±0.0012 Å, 7.8174±0.0025 Å and 7.5165±0.0016 Å, respectively, which should be compared with the values of 3.86769±0.00006 Å, 7.8143±0.0001 Å and 7.5125±0.0001 Å reported for the powder sample of ref. 17. The positional parameters obtained by the fitting are basically consistent with those in the article, though the fitting may not be fully satisfactory.

Then, the magnetic structures have been analyzed based on the crystal structure obtained above at 270 K and 250 K. (Although the increase of the peak intensity shown in Fig. 4(b) at the (0,2,0) point in the reciprocal space indicates the occurrence of the structural transition to the new phase at ~250 K with decreasing *T*, no effect is expected on the analyses of the magnetic structure at 250 K, because the order parameter of the new phase is, at least, negligibly small at the temperature.) The integrated intensities of the magnetic scattering were obtained by taking the differences between the integrated intensities at 300 K and those at 270 K or 250 K. (This automatically indicates that the superlattice reflections corresponding to the ~$3a_p$ (or ~$6a_p$) period modulation are not magnetic at least above 250 K.) The absorption- and Lorentz factor-correction were made. The magnetic structures have been determined by assuming that the magnitudes of the Co-moments and the absolute values of their cant angle are equal at the crystallographically equivalent positions. For the magnetic form factor of $Co^{3+}$, we used the isotropic values reported in ref. 19.

At 270 K, in order to distinguish which reflection, $0kl$ and $h0l$, is actually observed, accurate determination of the scattering angles of the magnetic reflections has been carried out by using the finer horizontal collimations. The results found around (0,1,1) or (1/2,0,1) and (0,2,0) or (1,0,0) points in the reciprocal space are shown in the top and bottom panels of Fig. 6, respectively. In the figure, the solid circles represent the intensity profiles of the nuclear reflections observed at 300 K and the open circles are the intensity profiles of the magnetic scattering at 270 K, which are extracted by subtracting the neutron count number at 300 K from that observed at 270 K at each scattering angle $2\theta$. It can be concluded from the figures that the 1/201 and 020 magnetic reflections have significant intensities, while the intensities of the 011 and 100 reflections are zero or negligibly small (note that $a<b/2$). Then, because the ferromagnetic magnetization is known to be in the $c$-plane from the result shown in Fig. 2, the direction of the ferromagnetic moment has to be along the $a$-axis. Magnetic reflections have also been observed at (0,1,0) and (0,3,0) points but not observed at (1/2,0,0) and (3/2,0,0). Based on these results, the magnetic structure finally obtained here is shown in Fig. 7. Figure 8 shows the integrated intensities of the magnetic reflections ($I_{obs}$) collected at 270 K against those of the model calculation ($I_{cal}$) at several reflection points. The results are also shown in Table I. The fitting is as good as that for the nuclear structure shown in Fig. 5, indicating that the obtained magnetic structure is essentially correct, even though we have ignored the $\sim 3a_p$ (or $\sim 6a_p$) modulation and just used the isotropic form factor of $Co^{3+}$. The result also indicates that the transition is purely magnetic or at least mainly magnetic. One of the characteristics of this magnetic structure is that all $Co^{3+}$ ions of $CoO_6$ octahedra are in the LS state and the magnetic moments of $Co^{3+}$ of $CoO_5$ pyramids have the coplanar but canted structure. The magnitude $\mu$ of the aligned moments of $Co^{3+}$ ions of the pyramids is 0.710($\pm$0.019) $\mu_B$. We think that $Co^{3+}$ is in the IS state, even though the ordered moment is much smaller than the value in the fully ordered state. The absolute value of the canting angle $\phi_1$ defined as shown in the bottom panel of Fig. 7 is 43.1($\pm$2.0)°. The sign of $\phi_1$ alternates along the $a$- and $c$-directions. The maximum value of the spontaneous magnetization in the $c$-plane, is expected to be 0.18 $\mu_B$/Co by the present fitting. (Due to the existence of the domain structure, the magnetization has the maximum value when $H$ is applied nearly along the [120] direction.) This result is consistent with the observed data shown in Fig. 3.

The magnetic structure obtained at 250 K (or in the antiferromagnetic phase) is shown in Fig. 9, where all $b$-planes formed of the $CoO_5$ pyramids have the same magnetic structure. Each $c$-plane has nonzero net magnetization and its direction rotates by 90°, when the $z$-position is shifted by $\sim c/2$, as shown by the arrows in the right part of the figure. Figure 10 shows the integrated intensities of the magnetic reflections collected at 250 K ($I_{obs}$) against those of the model calculation ($I_{cal}$) at several superlattice points. (No magnetic reflections appear at the fundamental Bragg points.) The results are also shown in Table II. The fitting is as good as that for the nuclear structure shown in Fig. 5, indicating that the obtained magnetic structure is essentially correct. The result also indicates that the transition is purely magnetic or at least mainly magnetic. The $Co^{3+}$ ions within the octahedra are in the LS state as 270 K and the magnitude of the aligned Co-moments within the pyramids is 1.090($\pm$0.031) $\mu_B$. The Co-moments cant by the angle $\phi_2 = 33.5(\pm1.5)°$ from the $a$- or $b$-axis. Then, each $CoO_2$ layer has the canted ferromagnetic structure similar to the structure of 270 K, in which the angle between the neighboring Co-moments along the $a$-axis is (180°-2$\phi_2$).

We have so far discussed the magnetic structure of $TbBaCo_2O_{5.5}$ at 270 K (in the ferromagnetic phase) and 250 K (just below the ferromagnetic-antiferromagnetic transition temperature) by analyzing the neutron diffraction data taken for a single crystal sample. Studies on the magnetic or crystal structures below $\sim$100 K and in the $T$-region between 100 K and 250 K, which have been found to be distinct from those of other phases in the present single crystal sample remain as the future problems.

## 4. Conclusion

The neutron diffraction intensities have been measured on the single crystal of $TbBaCo_2O_{5.5}$ and five transitions have been found. They are the metal to insulator transition (structural transition) at 340 K, to the ferromagnetic phase at 280 K, to the antiferromagnetic one at 260 K, and two unidentified transitions at $\sim$250 K and $\sim$100 K, with decreasing $T$. The magnetic structures have been extracted at $T$=270 K (ferromagnetic phase) and $T$=250 K (antiferromagnetic one). We have found that the $Co^{3+}$ ions of the $CoO_6$ octahedra are in the LS state, and those of the $CoO_5$ pyramids are possibly in the IS state at both temperatures. At $T$=270 K, the $Co^{3+}$-moments of the $CoO_5$ pyramids have the canted structure, where the direction of the ferromagnetic moment is along the $a$-axis. At $T$=250 K, they have the non-collinear structure, in which the direction of the net moment of the $CoO_2$ layer rotates by 90° when the $z$-position increases by $\sim c/2$.

Acknowledgments - Work at the JRR-3M was performed within the frame of JAERI Collaborative Research Program on Neutron Scattering. We would like to thank Dr. M. Matsuda and Mr. Y. Shimojo for the technical assistance at the spectrometer TAS-2.

Figure captions

Fig. 1     Schematic structure of $TbBaCo_2O_{5.5}$.

Fig. 2     Temperature dependence of the magnetizations taken for a sintered sample and as-cast and annealed crystals ($H$=1 T). Inset shows the temperature dependence of the electrical resistivity measured for a sintered sample.

Fig. 3     Magnetization of a single crystal of $TbBaCo_2O_{5.5}$ is shown at various temperatures against the magnetic field applied in the $c$-plane. In taking the data, the crystal orientation with respect to the $H$-direction was so chosen that $M$ at 270 K has the maximum value with the orientation-change within the condition of $H\perp$($c$-axis)

Fig. 4     (a) Fundamental or superlattice points of the unit cell with the size of $\sim a_p \times 2a_p \times 2a_p$ in the reciprocal space at 7 K. Solid circles represent the fundamental Bragg points. Other symbols are the superlattice points, where different symbols indicate the different temperature dependence of the intensities. (b) and (c) The peak intensities of several reflections are shown against $T$.

Fig. 5     Integrated intensities of the nuclear Bragg reflections collected at 300 K ($I_{obs}$) are plotted against those ($I_{cal}$) obtained by the fitting with the space group P*mmm*.

Fig. 6     Profiles of the 011 or 1/201 (top) and the 020 or 100/020 (bottom) reflections. Solid circles represent the reflection intensities at 300 K and the open circles show the magnetic scattering intensities at 270 K, which are extracted by subtracting the neutron count number at 300 K from that observed at 270 K at each scattering angle $2\theta$.

Fig. 7     Schematic magnetic structure of $TbBaCo_2O_{5.5}$ at 270 K(in the ferromagnetic phase). Top: The direction of the $Co^{3+}$-moments of the $CoO_5$ pyramids in the $b$-plane. Bottom: The magnetic structure of a $CoO_2$ layer.

Fig. 8     Integrated intensities of the magnetic reflections collected at 270 K ($I_{obs}$) are shown against those of the model calculation ($I_{cal}$).

Fig. 9     Schematic magnetic structure of $TbBaCo_2O_{5.5}$ at 250 K(in the antiferromagnetic phase). This structure has the period of $\sim 4a_p$ along the $c$-axis. The open arrows drawn in the right side of the figure show the directions of the net moments of the $CoO_2$ layers. All $Co^{3+}$ ions of $CoO_6$ octahedra are in the low spin state.

Fig. 10    Integrated intensities of the magnetic reflections collected at 250 K ($I_{obs}$) are plotted against those of the model calculation ($I_{cal}$).

Table I  Comparison of the observed intensities of the magnetic scattering with those calculated for the spin structure shown in Fig. 7.

$T$=270 K

| h k l | | $I_{\text{obs}}$ | $I_{\text{cal}}$* |
|---|---|---|---|
| 0 0 1 | | 0± 26 | 7 |
| 0 0 2 | | 340± 457 | 749 |
| 0 1 0 / | 1/2 0 0 | 823± 57 | 786 |
| 0 1 1 / | 1/2 0 1 | 605± 49 | 614 |
| 0 1 2 / | 1/2 0 2 | 248± 43 | 278 |
| 0 1 3 / | 1/2 0 3 | 402± 325 | 180 |
| 0 1 4 / | 1/2 0 4 | 138± 63 | 88 |
| 0 2 0 / | 1 0 0 | 401± 77 | 339 |
| 0 2 1 / | 1 0 1 | 238± 736 | 1 |
| 0 2 2 / | 1 0 2 | 897±1720 | 323 |
| 0 2 3 / | 1 0 3 | 149± 589 | 8 |
| 0 3 0 / | 3/2 0 0 | 174± 28 | 178 |
| 0 3 2 / | 3/2 0 2 | 126± 39 | 122 |
| 0 4 1 / | 2 0 1 | 85± 86 | <0.5 |
| 0 5 0 / | 5/2 0 0 | 103± 58 | 59 |

*for $\mu$=0.710$\mu_B$, $\phi_1$=43.1deg.

Table II   Comparison of the observed intensities of the magnetic scattering with those calculated for the spin structure shown in Fig. 9.

*T*=250 K

| h k l | | $I_{obs}$ | $I_{cal}$* |
|---|---|---|---|
| 0 0 1/2 | | 2733±343 | 2515 |
| 0 0 3/2 | | 561± 79 | 759 |
| 0 0 5/2 | | 349± 70 | 376 |
| 0 0 7/2 | | 225± 74 | 206 |
| 0 0 9/2 | | 148± 60 | 120 |
| 0 1 1/2 | / 1/2 0 1/2 | 1275±179 | 1141 |
| 0 1 3/2 | / 1/2 0 3/2 | 755± 78 | 891 |
| 0 1 5/2 | / 1/2 0 5/2 | 679± 77 | 540 |
| 0 1 7/2 | / 1/2 0 7/2 | 258± 70 | 319 |
| 0 2 1/2 | / 1 0 1/2 | 355± 69 | 271 |
| 0 2 3/2 | / 1 0 3/2 | 272± 64 | 260 |
| 0 2 5/2 | / 1 0 5/2 | 190± 61 | 199 |
| 0 3 1/2 | / 3/2 0 1/2 | 255± 56 | 247 |

*for $\mu$=1.090$\mu_B$, $\phi_2$=33.5deg.

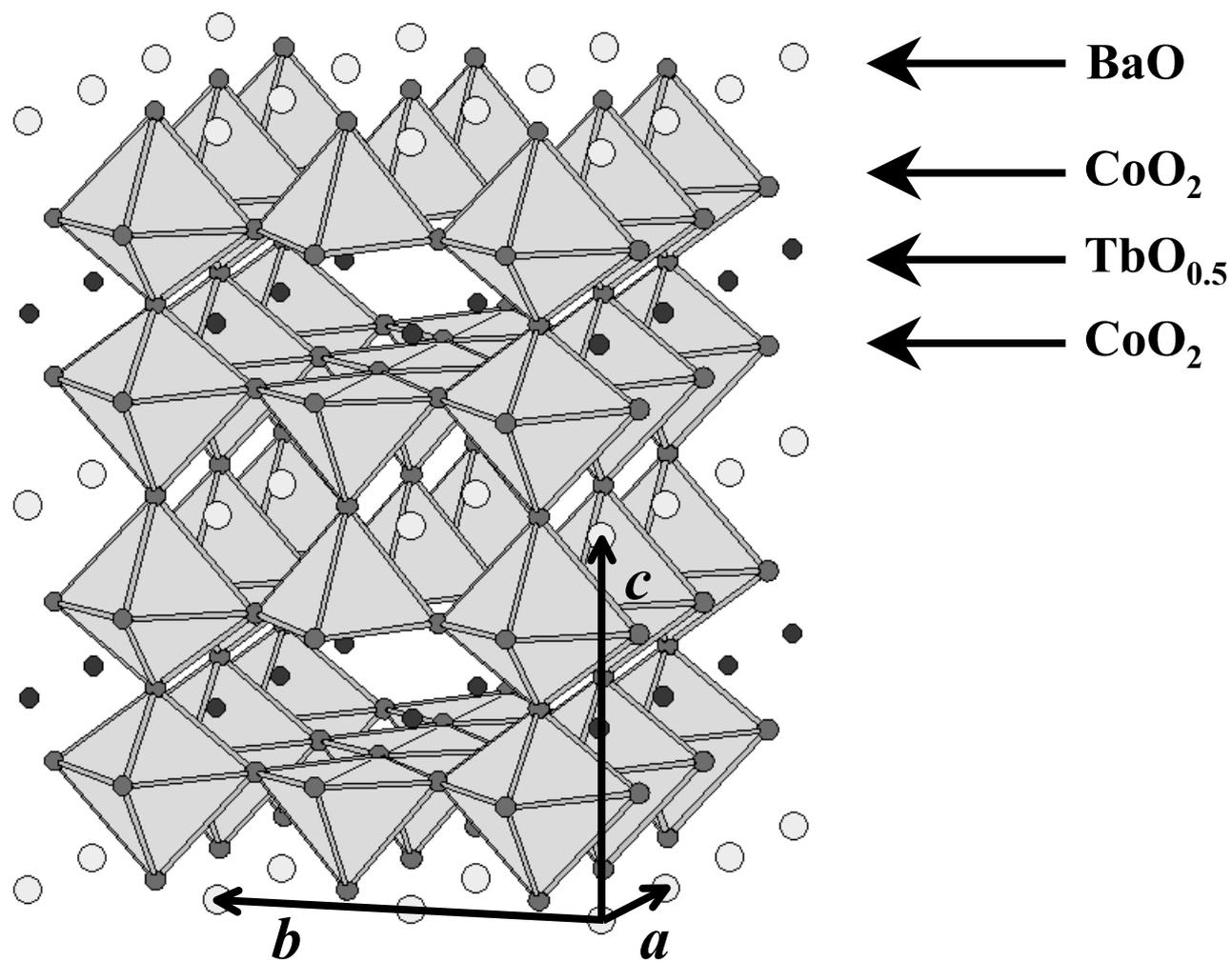

Fig. 1

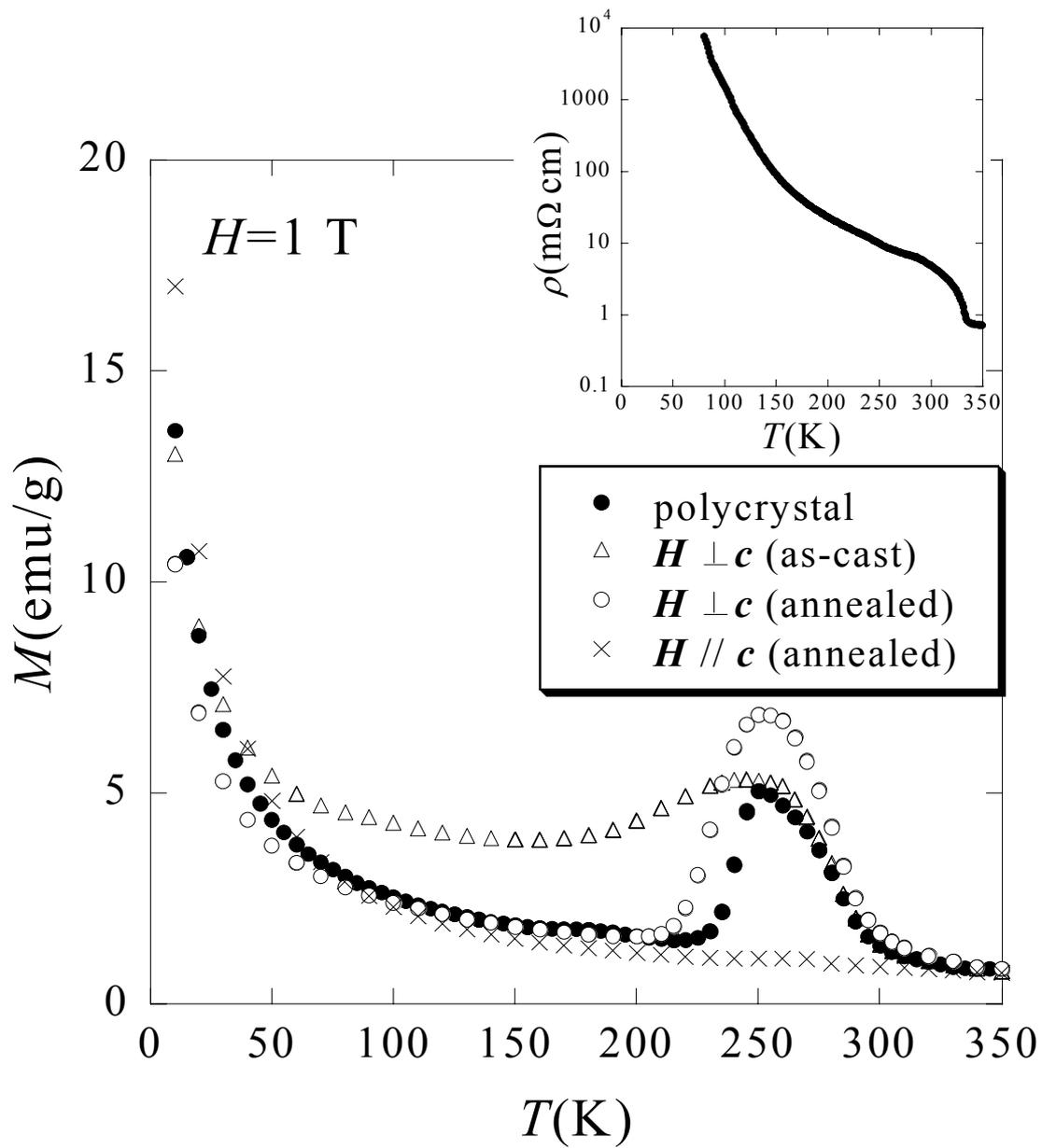

Fig. 2

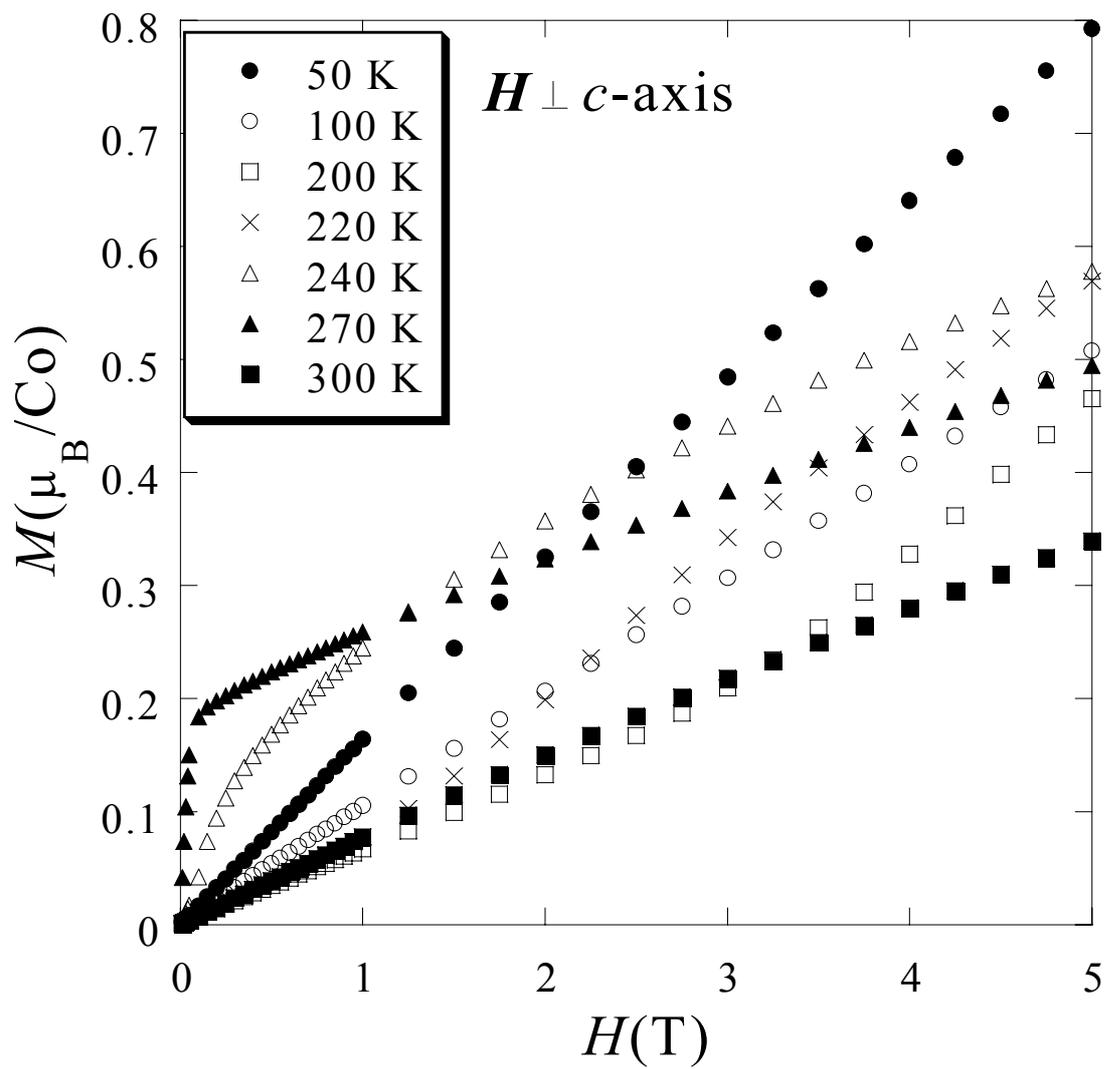

Fig. 3

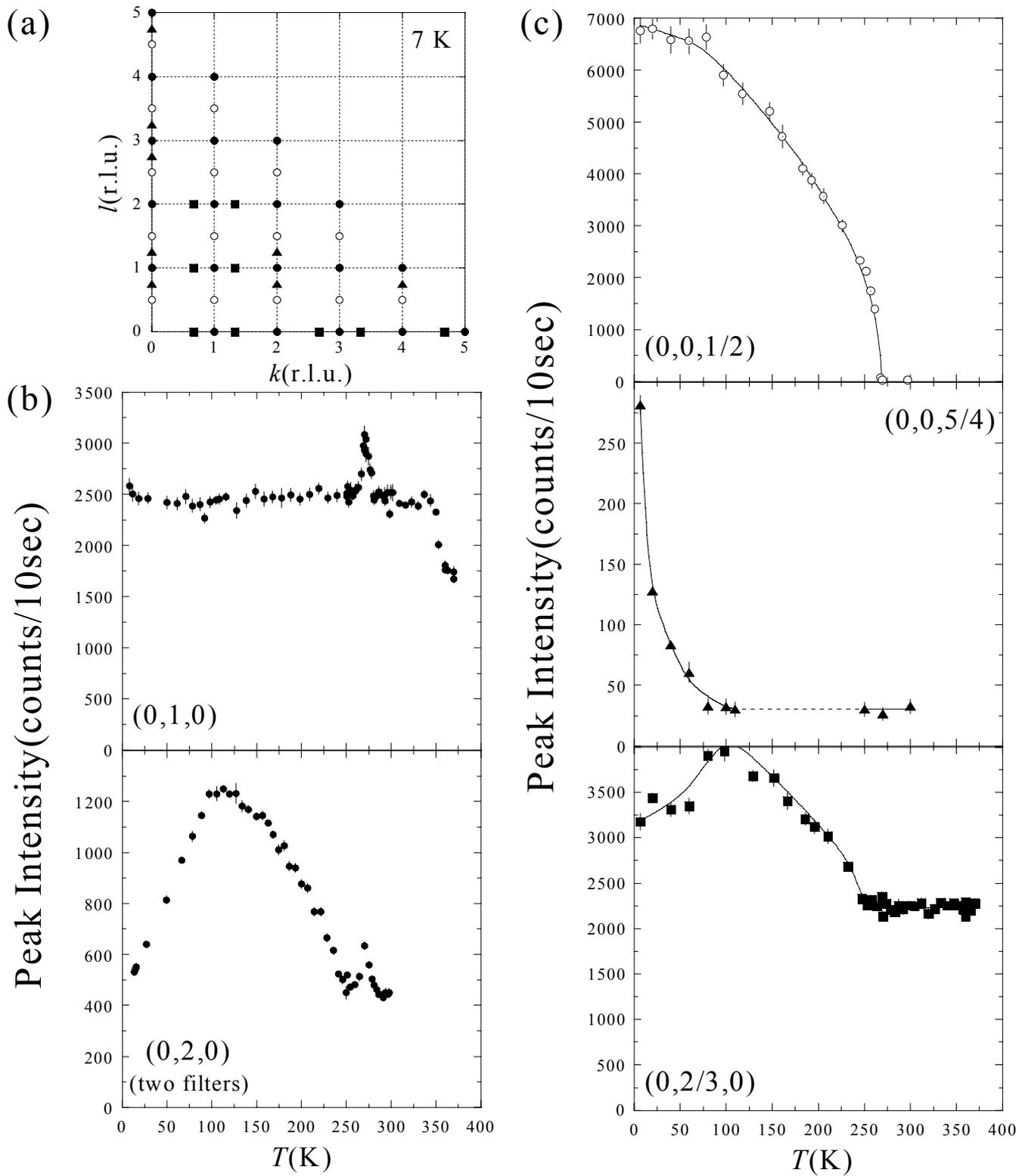

Fig. 4

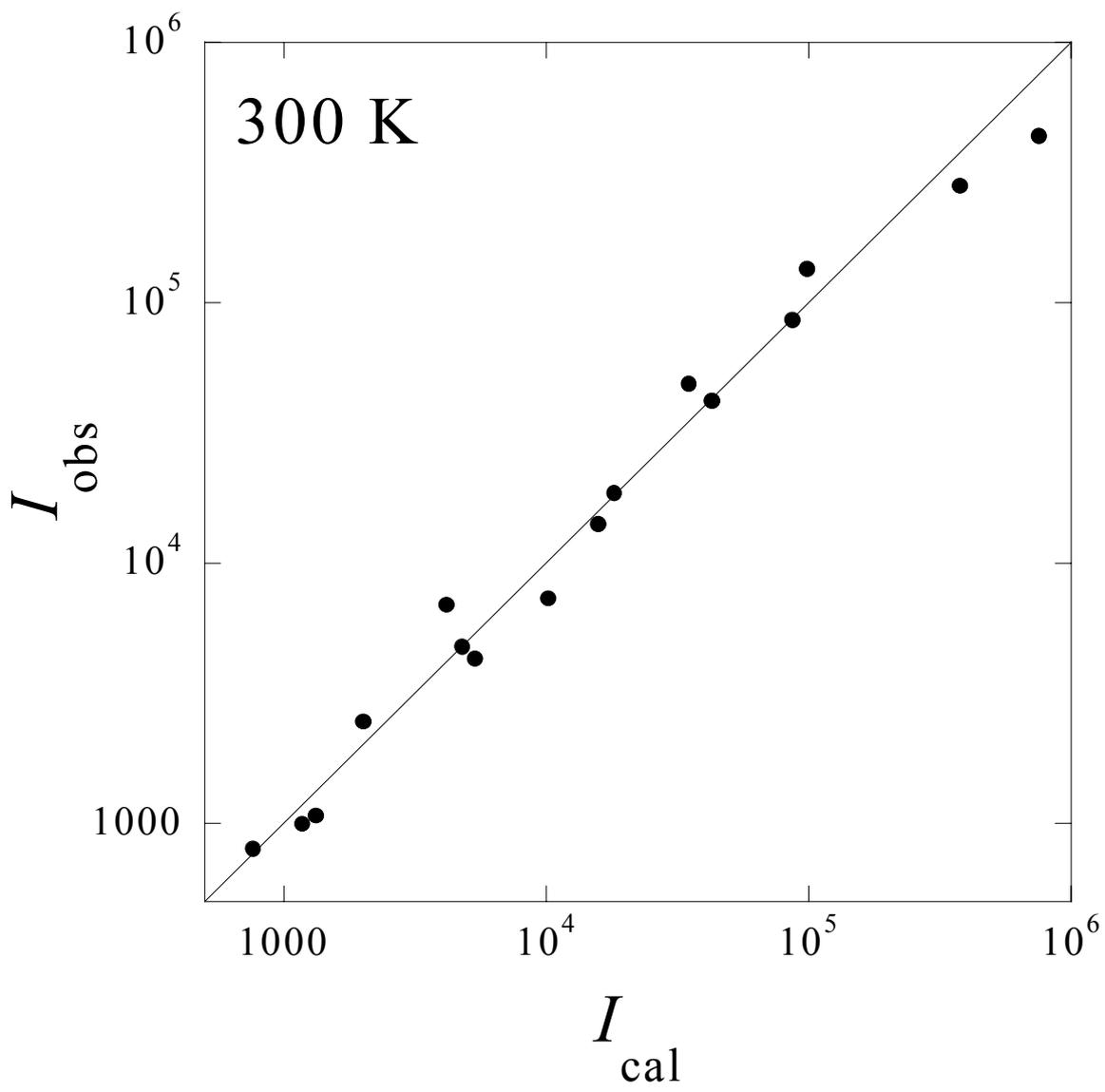

Fig. 5

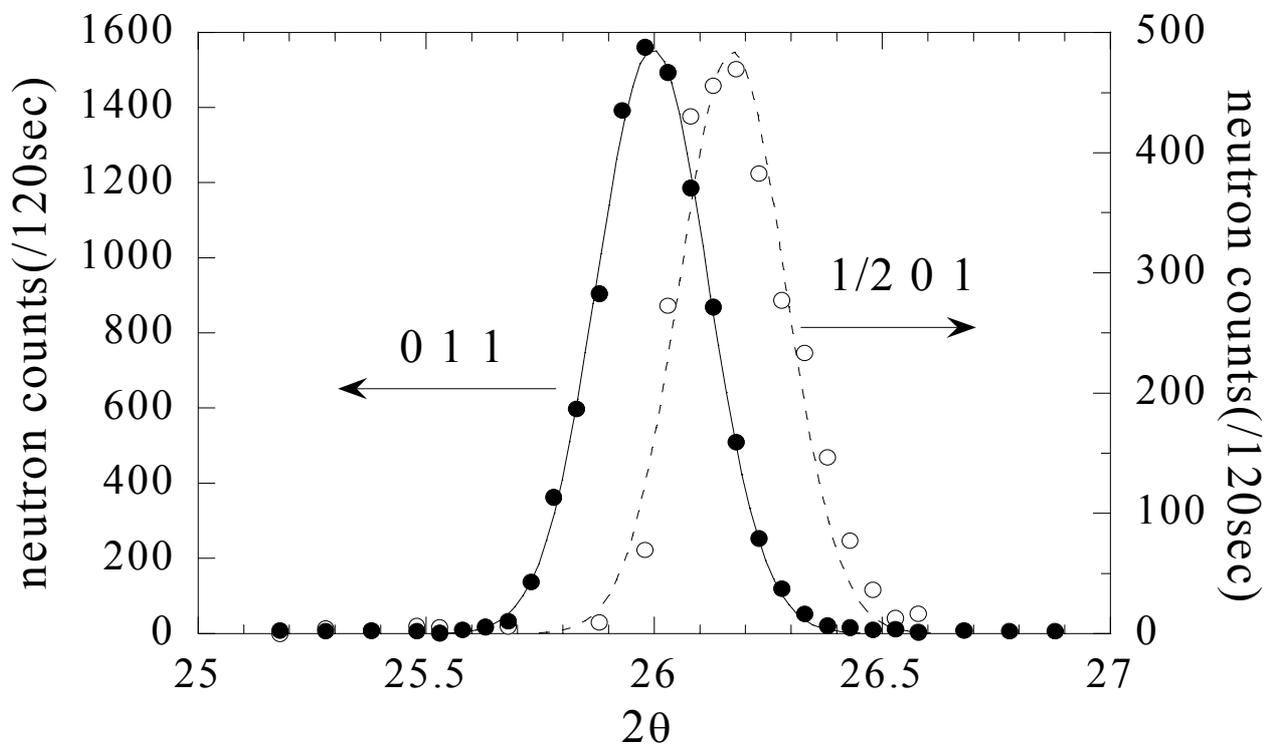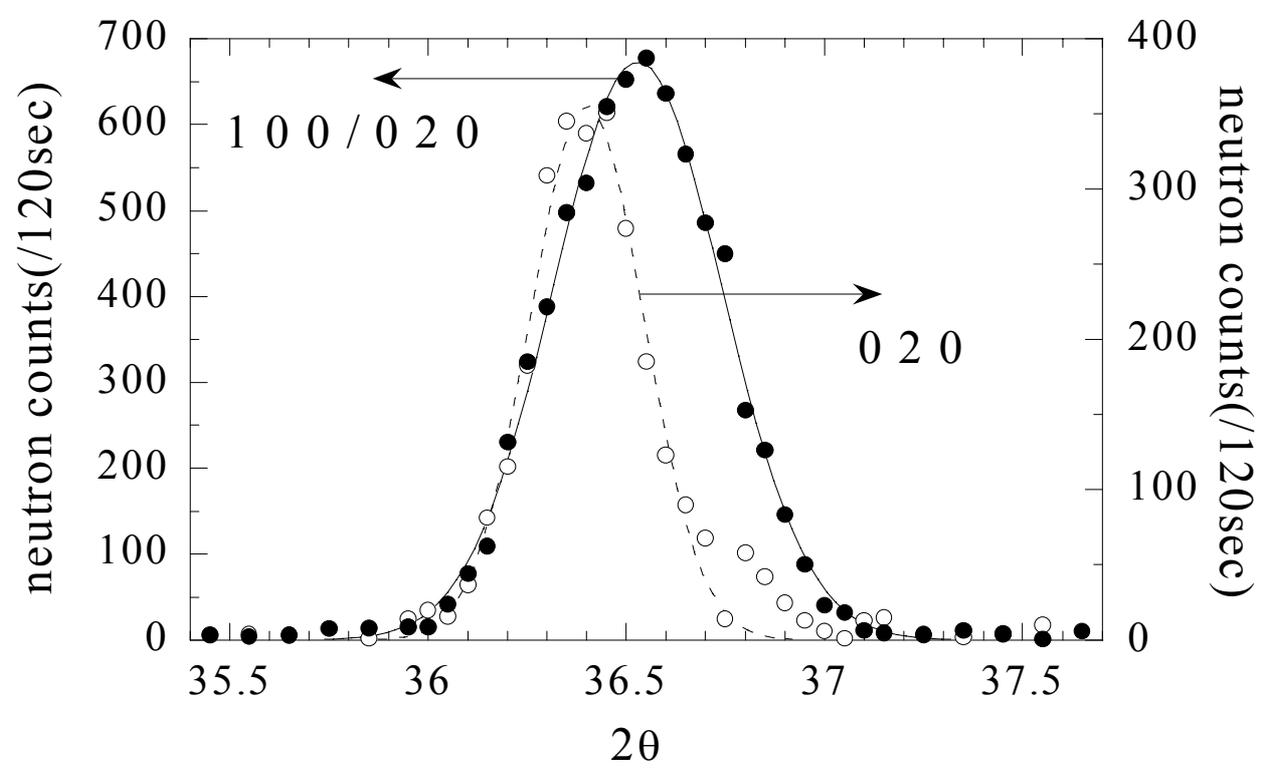

Fig. 6

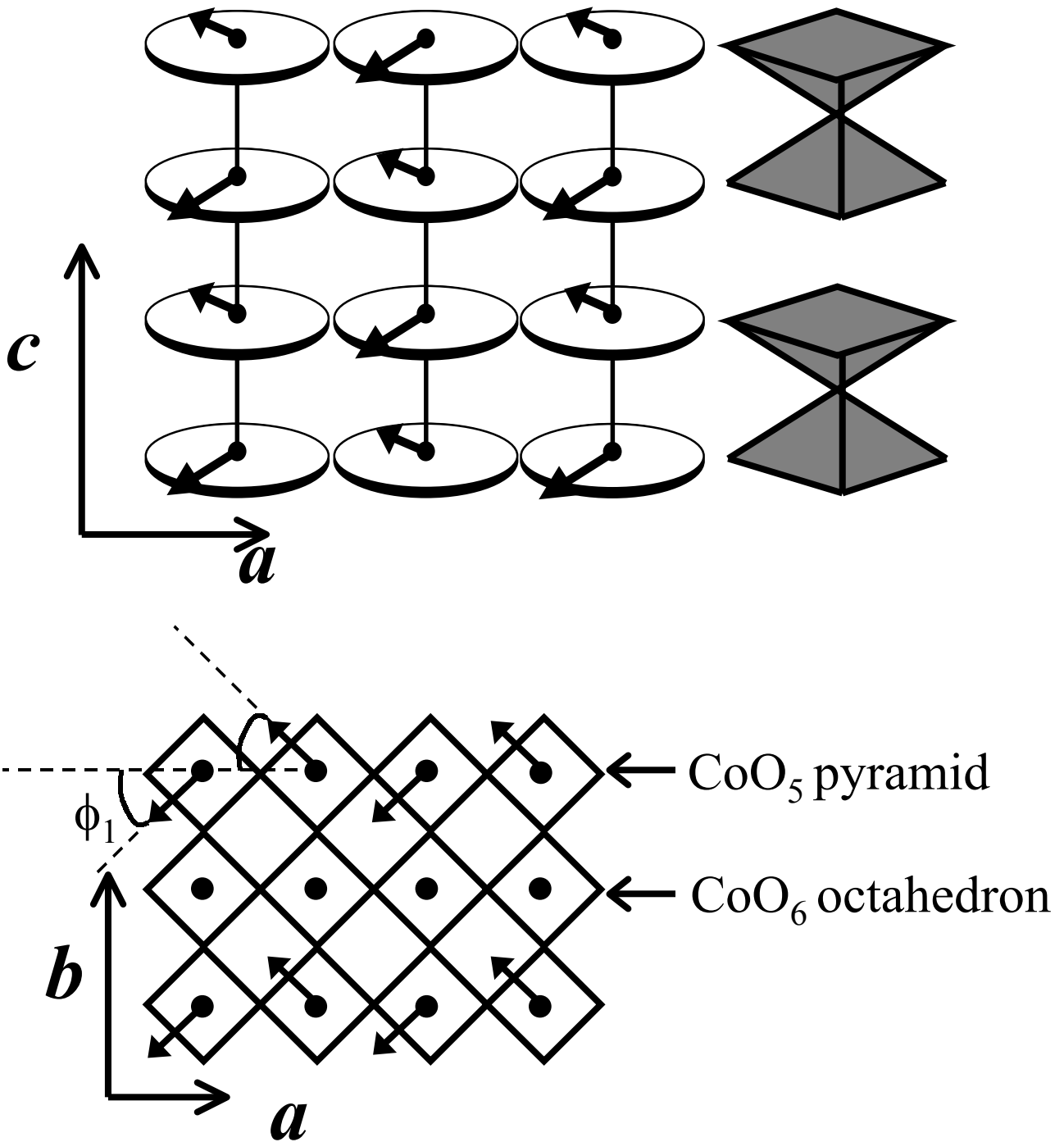

Fig. 7

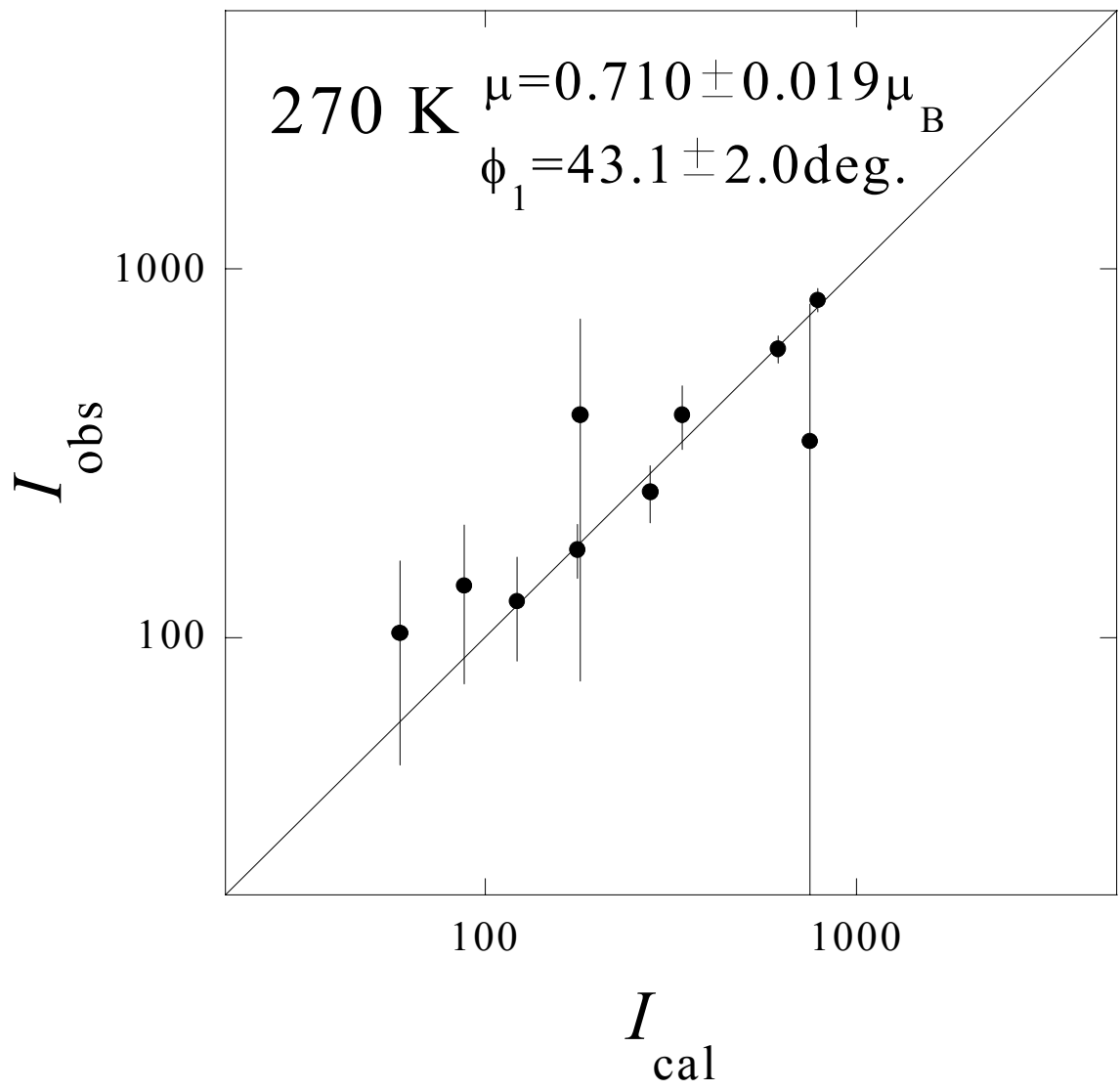

Fig. 8

250 K

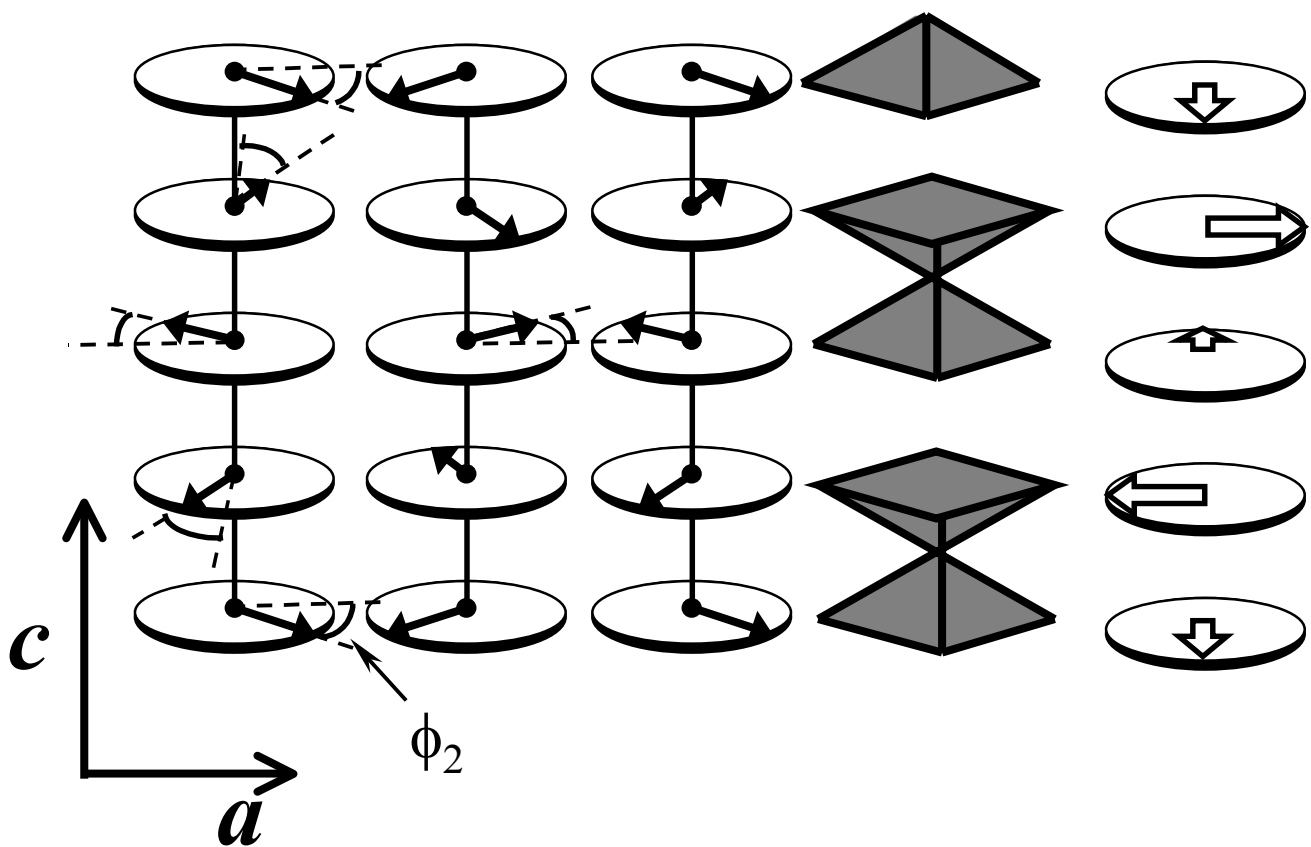

Fig. 9

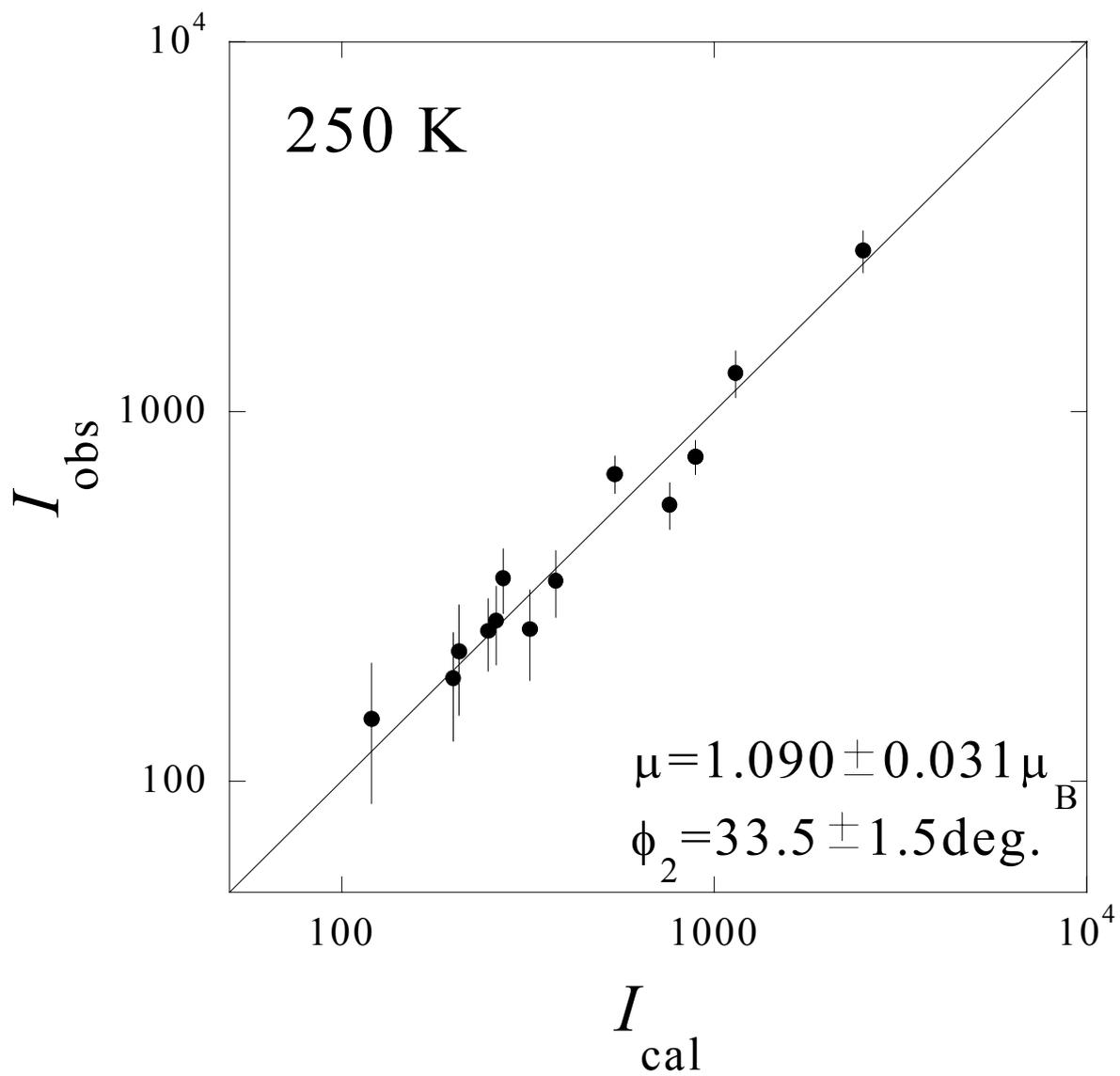

Fig. 10